\documentclass[iop]{emulateapj}

\usepackage{graphicx}
\usepackage{epstopdf}
\usepackage{subfigure}
\usepackage{pdfsync}
\usepackage{color}
\usepackage{natbib}
\font\fiverm=cmr5

\def\teq#1{$\, #1\,$}                         

\def\dover#1#2{\hbox{${{\displaystyle#1 \vphantom{(} }\over{
\displaystyle #2 \vphantom{(} }}$}}
{\catcode`\@=11                                                  
\gdef\SchlangeUnter#1#2{\lower2pt\vbox{\baselineskip 0pt\lineskip0pt    
\ialign{$\m@th#1\hfil##\hfil$\crcr#2\crcr\sim\crcr}}}}           
\def\gtrsim{\mathrel{\mathpalette\SchlangeUnter>}}               
\def\lesssim{\mathrel{\mathpalette\SchlangeUnter<}}

\def\gammat{\gamma_{\hbox{\fiverm T}}}

\newcommand{\vol}[2]{$\,$\rm #1\rm , #2.}

\shorttitle{Constraints on the Synchrotron Shock Model}
\shortauthors{J.~Michael Burgess et al.}

\begin{document}

\title{Constraints on the Synchrotron Shock Model for the Fermi GBM Gamma-Ray Burst 090820A}

              
\author{J. Michael Burgess,\altaffilmark{1}, 
Robert~D.~Preece\altaffilmark{1},
Matthew~G. Baring,\altaffilmark{2},
Michael~S.~Briggs\altaffilmark{1}, 
Valerie Connaughton,\altaffilmark{1},
Sylvain Guiriec,\altaffilmark{1} 
William S.~Paciesas\altaffilmark{1}, 
Charles A.~Meegan\altaffilmark{3}, 
P.~N. Bhat\altaffilmark{1}, 
Elisabetta Bissaldi\altaffilmark{4}, 
Vandiver Chaplin\altaffilmark{1}, 
Roland Diehl\altaffilmark{4}, 
Gerald~J.~Fishman\altaffilmark{5}, 
Gerard Fitzpatrick\altaffilmark{6}, 
Suzanne Foley\altaffilmark{6},
Melissa Gibby\altaffilmark{7}, 
Misty Giles\altaffilmark{7},
Adam Goldstein\altaffilmark{1}, 
Jochen Greiner\altaffilmark{4}, 
David Gruber\altaffilmark{4}, 
Alexander J.~van der Horst\altaffilmark{3}, 
Andreas von Kienlin\altaffilmark{4}, 
Marc Kippen\altaffilmark{8}, 
Chryssa Kouveliotou\altaffilmark{5}, 
Sheila McBreen\altaffilmark{6}, 
Arne Rau\altaffilmark{4}, 
Dave Tierney\altaffilmark{6}, and 
Colleen~Wilson-Hodge\altaffilmark{5}}              
              
\altaffiltext{1}{University of Alabama in Huntsville, 320 Sparkman Drive, Huntsville, AL 35899, USA}
\altaffiltext{2}{Department of Physics and Astronomy, MS 108,
      Rice University, Houston, TX 77251, U.S.A.  {\it Email: baring@rice.edu}}
\altaffiltext{3}{Universities Space Research Association, 320 Sparkman Drive, Huntsville, AL 35899, USA}
\altaffiltext{4}{Max-Planck-Institut f$\rm \ddot{u}$r extraterrestrische Physik (Giessenbachstrasse 1, 85748 Garching, Germany)}
\altaffiltext{5}{Space Science Office, VP62, NASA/Marshall Space Flight Center, Huntsville, AL 35812, USA}
\altaffiltext{6}{School of Physics, University College Dublin, Belfield, Stillorgan Road, Dublin 4, Ireland}
\altaffiltext{7}{Jacobs Technology}
\altaffiltext{8}{Los Alamos National Laboratory, PO Box 1663, Los Alamos, NM 87545, USA}              

\email{james.m.burgess@nasa.gov}

\begin{abstract}
Discerning the radiative dissipation mechanism for prompt emission in
Gamma-Ray Bursts (GRBs) requires detailed spectroscopic modeling that straddles
the $\nu F_{\nu}$ peak in the 100 keV - 1 MeV range.  Historically,
empirical fits such as the popular Band function have been employed with
considerable success in interpreting the observations. While extrapolations of the Band parameters can provide some physical insight into the emission mechanisms responsible for GRBs, these inferences do not provide a unique way of discerning between models. By fitting physical models directly this degeneracy can be broken,
eliminating the need for empirical functions; our analysis here offers
a first step in this direction. One of the oldest, and
leading, theoretical ideas for the production of the prompt signal is
the synchrotron shock model (SSM). Here we explore the
applicability of this model to a bright {\it Fermi} GBM burst with a
simple temporal structure, GRB~{\it 090820}A. Our investigation
implements, for the first time, thermal and non-thermal synchrotron
emissivities in the RMFIT forward-folding spectral analysis software
often used in GBM burst studies. We find that these synchrotron emissivities, together with a blackbody
shape, provide at least as good a match with the data as the Band GRB
spectral fitting function.  This success is achieved in both
time-integrated and time-resolved spectral fits.
\end{abstract}

\keywords{acceleration of particles --- gamma-ray bursts: individual (GRB 090820A) --- gamma rays: stars --- methods: data analysis --- radiation mechanisms: non-thermal --- radiation mechanisms: thermal}

\section{Introduction}

In the most popular paradigm for gamma-ray bursts of both long and short
durations, it is typically assumed that prompt $\gamma$-ray emission results from the
dissipation of kinetic energy in a relativistically expanding fireball
mediated by multiple internal shocks (e.g. see Piran 1999, or
M\'esz\'aros 2001, for reviews). These shocks are presumed to
diffusively accelerate a fraction of the electrons from thermal
upstream distributions to higher energies.
Usually only particles in the exponential tail 
of the Maxwellian are available for acceleration. 
Thus, for relativistic shocks, the expected outcome is that the particle 
distribution consists of a Maxwellian with a power-law tail at high energies.
Based on this scenario, the radiative emission should
consist of two components, quasi-thermal and non-thermal photons from
electrons spiralling along magnetic field lines in optically thin
regions of the jet. There could also be an additional photospheric
contribution of Planckian form, originating in distinct, optically thick
environs, perhaps interior to the regions spawning synchrotron emission.

To date, the characterization of GRB spectra has been dominated by the use of 
the empirical Band function \citep{Band:1993p20784}, a parametrized, smoothly broken power law
that was devised in the era of the BATSE experiment on the 
Compton Gamma-Ray Observatory (CGRO). Several 
authors have used measurements of the Band spectral shape parameters to infer properties of the 
physics involved in GRB emission. In particular, the fitted spectral indices defined by $N_{\gamma}\propto E^{-\alpha}$ below the \teq{\nu F_{\nu}} peak and $N_{\gamma}\propto E^{-\beta}$ above it, may be compared with 
values predicted from synchrotron emission: the low-energy self-absorption index, $\alpha$, of $+1$ (in 
photon flux units), the synchrotron `line of death'  index of $-2/3$, the `second line of death' 
at the fast cooling  
value of $-3/2$, the high energy index, $\beta$, characterizing power-law particle acceleration, and the 
various spectral differences between these 
\citep{Preece:1998p11011,Lloyd:2000p23470,LloydRonning:2002p17917,Preece:2002p11128}. 
However, it becomes difficult to discern between models through the Band function when the fitted 
low energy indices represent a power law only asymptotically, and when  many models predict 
similar Band indices. In fact, the Band function's inherent shape and curvature only loosely
approximates the shape of the applicable physical models making it difficult to draw conclusions
about emission mechanisms directly from Band function fits. A way to break this degeneracy is by
fitting more realistic emission models to the data, which in addition provides deeper insights into the burst
environment. In Section \ref{sec:model} we detail the emission model that we use to fit GRB spectra.
We present this model as a first step. In future work we will explore additional models in an attempt
to discern between them. We describe our observational results with this model in Section
\ref{sec:observe}.


\section{Model and Motivation}
\label{sec:model}

We propose to test an emission model composed of synchrotron emission and a thermal blackbody. This model is the most general form of the standard fireball model.
Non-thermal synchrotron emission is historically the most
favored process invoked to explain prompt GRB signals.
The motivation for the inclusion of a blackbody component comes not only from theory \citep{Goodman:1986,Meszaros:2002,Rees:2005} and previous searches \citep{Ryde:2009}  but also the recent discovery of a significant quasi-thermal component in GRB spectra \citep{Guiriec:2010p26991,Ryde:2010p21261}. However, in 
Guiriec et al. (2010; GRB~{\it 100724}B) as well as Ryde et al. (2010; GRB~{\it 090902B}), 
the non-thermal portion of the spectra is approximated by empirical functions that lack direct associations with the physical parameters.

In order to model optically-thin synchrotron emission in a physical way, we 
adopt the parametrization presented in Baring \& Braby (2004, hereafter BB04), which was 
modified slightly from the choice of Tavani (1996).
Theory and numerical simulations predict that the electron energy distribution 
resulting from diffusive shock acceleration should be composed of 
two components (e.g. see Baring 2011 for an overview), which to 
first order can be approximated by a superposition of 
a relativistic Maxwellian and a super-thermal power-law tail:
\begin{equation}
   n_e(\gamma )\; =\; n_{0} \biggl\lbrack\;
    \Bigl( \dover{\gamma}{\gammat} \Bigr)^2\,
   e^{-\gamma/\gammat } + \epsilon \,
    \Bigl( \dover{\gamma}{\gammat} \Bigr)^{-\delta}\,
    \Theta \Bigl( \dover{\gamma}{\eta\gammat} \Bigr)\, \biggr\rbrack\, ,
 \label{eq:elec_dist}
\end{equation} 
where \teq{\Theta (x)} is a step function with \teq{\Theta (x)=1} 
for \teq{x\geq 1} and zero otherwise, and \teq{\gammat} 
is a measure of the post-shock electron temperature.  
This is a quasi-isotropic distribution, in the comoving frame of
reference of the GRB outflow (the mildly-relativistic speed of an internal 
shock in this frame does not change this form significantly), 
with the dependence 
on pitch angle being omitted for simplicity, though it can be 
incorporated in the \teq{n_0} factor.  The shock acceleration 
electron distribution therefore depends on five
parameters, three of which, the power-law index, \teq{\delta}, 
the relative normalization, \teq{\epsilon} (which can be related to the acceleration efficiency), 
and the product \teq{\eta\gammat}, which 
defines the minimum Lorentz factor of the power-law, pass unmodified into the expression of the photon flux
and are thus fit parameters for the GRB data. 
In Tavani's original exposition \teq{\eta} was fixed to unity and the power law 
component smoothly joined to the exponential portion of the Maxwellian (i.e., with virtually no discontinuity).
This would be the case of `saturated' acceleration, where all of the
electrons above the peak in the Maxwellian have been accelerated. BB04 indicated that values
\teq{\eta\sim 3} and \teq{\epsilon\lesssim 0.1}  closely reflect
populations usually found in simulations of shock acceleration even ones based on
diverse and contrasting approaches (e.g. see Niemiec \& Ostrowski 2004; Spitkovsky 2008;
Baring 2011; and references therein). For
simplicity and general facility of spectral fitting, we adopt the
compact form in Eq.~(\ref{eq:elec_dist}), deferring
direct fitting with specific simulation model output to future studies. 
Here, \teq{\eta =3} is adopted as a representative value that incurs no
significant discontinuity in transitioning from the Maxwellian to the 
non-thermal population when \teq{\epsilon\lesssim 0.1}.

In a truly physical model, 
the electron distribution function should be perfectly continuous, contrasting Eq.~(\ref{eq:elec_dist}). Here, we have left both 
\teq{\eta} and \teq{\epsilon} as parameters free to vary, observing that folding the distribution with the synchrotron 
emissivity function in Eq.~(\ref{eq:synch_flux}) below 
yields continuous emission spectra.  Thus, while not explicitly joining 
the two components of Eq.~(\ref{eq:elec_dist}) smoothly, the subsequent fitting of GRB spectral data 
provides a robust and informative indication concerning the relative contribution of each 
component, as was done in BB04.  More precise modeling with truly continuous 
electron distributions is left for future investigations, but is unlikely to alter the essential 
conclusions of our work here.
 
To determine the radiation {\it flux}, \teq{F_{\nu}(\mathcal{E})\propto\mathcal{E}N_{\gamma} }, emitted by these electrons, this distribution 
is convolved with the standard synchrotron emissivity 
(e.g. Rybicki and Lightman, 1979; see also BB04):
\begin{equation}
   F_{\nu}(\mathcal{E})\; \propto\; \int_1^{\infty} n_e(\gamma ) \, 
   \mathcal{F} \left( \dover{\mathcal{E}}{\mathcal{E}_c}\right) \, d\gamma\quad ,
 \label{eq:synch_flux}
\end{equation}
where
\begin{equation}
   \mathcal{F}\left(w\right) \; =\; w \int_w^{\infty } K_{5/3}(x) \, dx
 \label{eq:synch_func}
\end{equation}
expresses the single-particle 
synchrotron emissivity (i.e. energy per unit time per unit volume)
in dimensionless functional form.
The characteristic scale for the synchrotron photon energy is
\begin{equation}
   \mathcal{E}_c\; =\;\dover{3}{2}\; \dover{B}{B_{\rm cr}}\,\Gamma\, \gamma^2\, m_ec^2\quad ,
  \label{eq:ergcsynch}
\end{equation}
where \teq{B_{\rm cr} = 4.41 \times 10^{13}}Gauss is the quantum 
critical field. 
When convolved with the distribution in Eq.~(\ref{eq:elec_dist}), 
the substitution \teq{\gamma\to \eta\gammat} in Eq.~(\ref{eq:ergcsynch})
then defines the scale for the break energy of the synchrotron continuum
resulting from the truncated power-law portion of the distribution 
(see Table~1).
 In modeling prompt burst emission, the relativistic 
nature of the outflow introduces an extra parameter, the bulk 
Lorentz factor \teq{\Gamma} of the flow, which blueshifts the 
spectrum so as to introduce the \teq{\Gamma} factor in 
Eq.~(\ref{eq:ergcsynch}), so that
Eq.~(\ref{eq:synch_flux}) then expresses the synchrotron 
flux in the observer's frame.  Accordingly, while the electron 
distribution parameters \teq{\delta} and \teq{\epsilon}
can be constrained by prompt emission spectroscopy,
the precise values of \teq{\gammat}
and the environmental quantities \teq{B} and \teq{\Gamma} are
indeterminate, being subsumed in the single parameter 
\teq{\Gamma \eta^2\gammat^2 B} that is defined by a spectral fit
in a given time interval.

For fits where non-thermal synchrotron components dominate, 
the energy of the \teq{\nu F_{\nu}} peak determines the value of 
the  peak energy.  Well below this
structure the flux index is \teq{+1/3} and well above it, the flux index 
is \teq{-(\delta -1)/2}.  This is the simplest synchrotron model to consider. 
Strong cooling synchrotron models possess a similar mathematical 
character, but elicit a gentler break and a steeper spectrum below the 
break that is often more difficult to fit to observations. Treatment 
of such cooling models, and inverse Compton scenarios will be 
deferred to future work. We note also that models where 
\teq{\epsilon\ll 1} and
the non-thermal synchrotron component is small or insignificant,
the high energy tail of the thermal synchrotron component is 
necessarily exponentially declining with energy.  Such forms 
have severe difficulty in fitting GRB spectra that possess
extended power-law tails, a common occurrence, yielding 
\teq{\epsilon \gtrsim 0.1} as an anticipated frequent inference
in this GRB spectroscopy protocol.

In summation, our emission model consists of a 
two-component synchrotron function (thermal and power-law), 
plus a blackbody, all boosted from the outflow frame, by the bulk Lorentz factor \teq{\Gamma}, to the
observer's frame.  Along with the blackbody component, this spectral model has seven fit parameters;
values for two of these parameters, \teq{\eta} and \teq{\delta}, are fixed for reasons detailed in
Section ~\ref{sec:observe}. Owing to the intensive numerical integration involved, such functions
have previously not been used for forward-folding spectral fitting, particularly in the 
CGRO/BATSE era.
We have implemented this photon model into the RMFIT spectral analysis software
and demonstrate our technique by fitting fitting the prompt emission of GRB~{\it 090820}A; one of the
brightest GBM bursts with simple temporal structure.

\section{Observations}
 \label{sec:observe}

On 20 August 2009, at T$_\mathrm{0}$=00:38:16.19 UT, the Gamma-Ray Burst Monitor (GBM)
onboard the Fermi Gamma-ray Space Telescope triggered on the very
bright GRB~{\it 090820}A~\citep{GCN-GBM}. This GRB also triggered 
Coronas Photon-RT-2~\citep{GCN-CORONAS}. The burst location was initially not in the FOV of the Large Area Telescope (LAT) onboard Fermi but was bright enough to result in a Fermi spacecraft
repointing maneuver. However, Earth avoidance constraints prevented such a maneuver
until 3100 sec after the burst trigger and the burst was not detected at higher energies by the LAT. 
The most precise position for the
direction of the burst comes from the GBM trigger data which localizes the burst to a
patch of sky centered on RA = 87.7 degree and Dec = 27.0 degree (J2000) with a 4 degree error, statistical and systematic. The current best model for systematic errors is 2.8 degrees with 70\% weight and 8.4 degrees with 30\% weight \citep{briggs_ann}. We verified that our analysis does not change significantly using instrument response functions for assumed source locations throughout this region of uncertainty. 

 GBM is composed of 12 sodium iodide (NaI) detectors covering an energy range from 8 keV to 1 MeV and two bismuth germanate (BGO) detectors sensitive between
200 keV and 40 MeV~\citep{Meegan:2009p28139}. Figure~\ref{figure1} (top two panels) shows the
light curve of GRB~{\it 090820}A as seen by GBM, 
from 8 to 200 keV in the NaI detectors (top) and from 200 keV to 40 MeV in the BGO detector (bottom). GBM triggered on a weak precursor which we do not include in the analysis. The main light curve begins at T$_{0}$ + 28.1s.
The main structure of the light curve consists of a fast rising pulse with an exponential decay lasting until T$_{0}$+60 s. A second, less intense, peak beginning at T$_{0}$+30 s is superimposed on the main peak. 
With such a high intensity and simple structure, this GRB allows for
detailed time-resolved spectroscopy. Because this burst is intense, calibration issues make the Iodine K-edge (33 keV)  prominent
in the count spectra owing to small statistical uncertainties, and we remove energy channels contributing to this feature from our spectral
fits. In addition, an effective area correction is
applied between each of the NaI detectors and the BGO 0 during the fit
process. This correction of $\approx$~23\% is used to account for
possible imperfections in the response models of the two detector types.

We simultaneously fit the spectral data of the NaI detectors with a
source angle less than 60 degrees (NaI  1 and 5) and the data from the
brightest BGO detector (BGO 0) using the analysis package
RMFIT. We use a forward-folding technique that
convolves the detectors' response with the proposed photon model to
generate a count spectrum to compare to the data; the parameters of the photon model are then adjusted so as to optimize the Castor C-stat statistic. The Castor C-stat differs from Poisson likelihood by an offset which is a constant for a particular dataset.

 We perform a fit to the integrated spectrum and find that it is best
represented by synchrotron emission from thermal and power-law
distributed electrons with an additional blackbody component characterized by a kT~$\approx$~42 keV (C-Stat/DOF = 558/353). The \teq{\nu F_{\nu}} spectrum is displayed in Figure \ref{figure2} and the best-fit values in Table \ref{table1}. We also performed a fit using the Band function (C-Stat/DOF = 593/355). We find in concordance with BB04 that emission from
power-law synchrotron dwarfs the emission from thermal synchrotron by at
least 3 orders of magnitude. The value of $\eta$ is fixed to 3, 
the choice adopted by BB04: it is a value that accommodates
distributions typically determined by shock acceleration simulations.
When fitting the power-law synchrotron component
we have to fix the value of the power-law index to its
best fit value to remove a correlation between the amplitude and the
index; this does not change the fit statistic but does mean that the
amplitudes obtained are valid only for that index. The inferred electron distribution from this fit is shown in Figure \ref{figure3}. We note
that the inability to simultaneously constrain the power-law index and 
amplitude of the synchrotron function may be solved in future studies by
including joint fits with LAT data, whenever available.

For the time resolved analysis we fit four bins labeled {\textbf a}, {\textbf b}, {\textbf c} and {\textbf d} as shown in Figure \ref{figure1} with the various synchrotron models. The corresponding electron distributions inferred from these fits are displayed in Figure \ref{figure5}. We also fit the Band function to each spectrum to show that in nearly all  cases the physical models can fit the data as well as the Band function. We chose the time binning by finding a balance between high signal-to-noise and evolution of the spectral shape so that we can identify the time evolution of each component throughout the burst. Where possible, we fit all three components simultaneously. Due to the similarity in the spectral shapes of the low energy portions of the thermal synchrotron and power-law synchrotron components it is not always possible to constrain all of the fit parameters especially when one component is much stronger than the other. Therefore, when one component is dominant we include only that component in the fit. The ability to fit both components in the time integrated fit is most likely due to the fact that both components are significant over the interval.

From bins {\textbf b} to {\textbf c} the spectrum is best described by  
synchrotron emission from power-law distributed electrons in addition  
to a blackbody (Table \ref{table1} and Figure~\ref{figure4}). The thermal synchrotron component is too weak to  
meaningfully include it in the fit. We find that the intensity of the  
power-law synchrotron increases significantly from bin {\textbf b} to {\textbf  
c} while the blackbody component remains nearly constant in intensity.  
The spectral
index of the electrons in these intervals varies from -4.4
to -5.9.
Such values are consistent with those expected from diffusive
acceleration theory, for the specific case of superluminal shocks
\citep{Baring:2010p26599}, i.e. those where the mean magnetic
field angle to the shock normal is significant. This geometrical  
requirement
establishes efficient convection of particles downstream of  
relativistic shocks,
thereby steepening their acceleration distribution. 
The blackbody component decreases in intensity at this  
point but the temperature remains constant within errors.
In bins {\textbf a} and {\textbf d}, with weaker emission, several models are essentially statistically tied.
It is possible that PLS+BB persists throughout the entire GRB. Alternatively, the GRB
could even begin in bin {\textbf a} with thermal synchrotron emission and transition to
the PLS+BB emission. If this were true we would be seeing emission from electrons that have not yet been accelerated into a power-law distribution by the shock. The C-stat values for all of the models fit in each bin are displayed in Table \ref{table2}.

While it is not possible to constrain all parameters in all the bins, it
should be stressed that this is due to natural correlations in the
synchrotron functions. These difficulties do not arise when using the
Band function because it has a simpler parametrization.

\section{Discussion and Conclusions}

In this paper, we have shown that thermal and non-thermal synchrotron
photon models, with an additional blackbody, are well consistent with the
emission spectra of GRB {\it 090820}A in various time intervals. These are physical models that afford the ability to constrain
parameters that are physically meaningful, for example key descriptors
of the electron distribution that is motivated by shock acceleration
theory. By implementing these models into a forward-folding spectral analysis software we have been able to directly constrain many of the physical model parameters and their respective errors; a first in the field of GRB spectroscopy. This constitutes
substantial progress over the use of the empirical Band function to fit
prompt GRB spectra, which has been a nearly universal practice to date. 
The results presented here enable more rigorous statements about the
validity of GRB emission models, moving the study of prompt burst
emission into a new era.

Our modeling has focused on the standard synchrotron shock model with
the addition of a blackbody component. The spectral fitting reveals a
complex temporal evolution of the separate components. While spectral
evolution is a well-known feature of GRBs, this type of fitting can enable \textit{direct} physical
interpretation of the evolution. These fits provide evidence that the line of death issue (Preece et al. 1998, 2002)
can be overcome naturally with a combination of synchrotron and
blackbody emission: the prominence of a blackbody component with its
flat Rayleigh-Jeans portion would derive a comparably-fitted Band
function with a flat low-energy index. This was also suggested by Guiriec et al. (2010) where the authors used simultaneous fits of the Band function and a blackbody. Note that it is possible that
other physical models may, in fact, produce superior fits to the data
for GRB~{\it 090820}A and other bursts. Strongly-cooled synchrotron
emission, inverse Compton and jitter radiation are popular candidates,
and our work here motivates the future development of RMFIT 
software modules for these processes.


A principal finding of the analysis in this paper is that the power-law
synchrotron component is orders of magnitude more intense than the
thermal synchrotron component during the peak of the burst, the latter
contributing at most a few percent of the flux. This confirms the
finding of BB04 for BATSE/EGRET bursts GRB {\it 910503}, 
GRB {\it 910601} and GRB {\it 910814}, which was a theoretically-based
perspective that did not fold models through the detector response
matrices. They had noted that full plasma and Monte Carlo diffusion
simulations of shock acceleration clearly predict a power-law tail in
the particle distribution that smoothly extends from the dominant
thermal population (e.g. see also Baring {2011}, and references therein).
This tail is several orders of magnitude smaller than what is found when 
fitting synchrotron emission to burst spectra. It is not clear how such
non-thermally-dominated distributions can arise near shocks, providing a
conundrum for the standard synchrotron shock model. Limited 
smoothing of the sharp peak of the non-thermal electron component 
will not alter this conclusion.

This result is also in accord with Guiriec et al. (2010), in their
analysis of GRB~{\it 100724}B, who fitted its GBM spectra with a
combination of the Band model and a blackbody. They too found that an
unrealistically high efficiency for the acceleration mechanism or 
a source size smaller than the innermost stable orbit of
a black hole was required to invoke the standard fireball model for
explaining the origin of the $\gamma$-ray emission. Therefore, it was
surmised therein that the outflow from the jet was at least partially
magnetized.

To conclude, the success of this analysis in isolating the 
relative contributions of a handful of distinct spectral components indicates that it is
imperative for the field of GRB spectroscopy to move away from the use
of the empirical fitting functions: many physical models can asymptotically approximate
the Band spectral indices, rendering it difficult to discern between
them particularly near
the $\nu F_{\nu}$ peak. Instead, direct comparisons of the fitted physical models are
possible, and are required to truly discriminate between the various
emission processes. The fitting of physical SSM/blackbody spectra here
offers a clear advance beyond empirical fits, and provides the impetus
for further development and deployment of physical modeling of prompt
burst emission spectra.

\acknowledgements{We thank the referee for many useful
comments that helped clarify the presentation. JMB thankfully acknowledges the support of the Alabama Space Grant Consortium through 
NASA Training Grant NNX10AJ80H. MGB is grateful for support under NASA's Fermi Guest Investigator
program, Cycle 2, through grant NNX09AT80G. AJvdH was supported by NASA grant NNH07ZDA001-GLAST.

\def\mn{M.N.R.A.S.}
\def\aas{{Astron. Astrophys.}}
\def\aassupp{{Astron. Astrophys. Supp.}}
\def\apss{{Astr. Space Sci.}}
\def\apj{ApJ}
\def\apjl{ApJ}
\def\nat{Nature}
\def\aaps{{Astron. \& Astr. Supp.}}
\def\aa{{A\&A}}
\def\apjs{{ApJS}}
\def\sp{{Solar Phys.}}
\def\jgr{{J. Geophys. Res.}}
\def\grl{{Geophys. Res. Lett.}}
\def\jphysb{{J. Phys. B}}
\def\ssr{{Space Science Rev.}}
\def\araa{{Ann. Rev. Astron. Astrophys.}}
\def\nature{{Nature}}
\def\asr{{Adv. Space. Res.}}
\def\prc{{Phys. Rev. C}}
\def\prd{{Phys. Rev. D}}
\def\pr{{Phys. Rev.}}

\clearpage

\begin{figure}
\includegraphics[scale=1, angle=0]{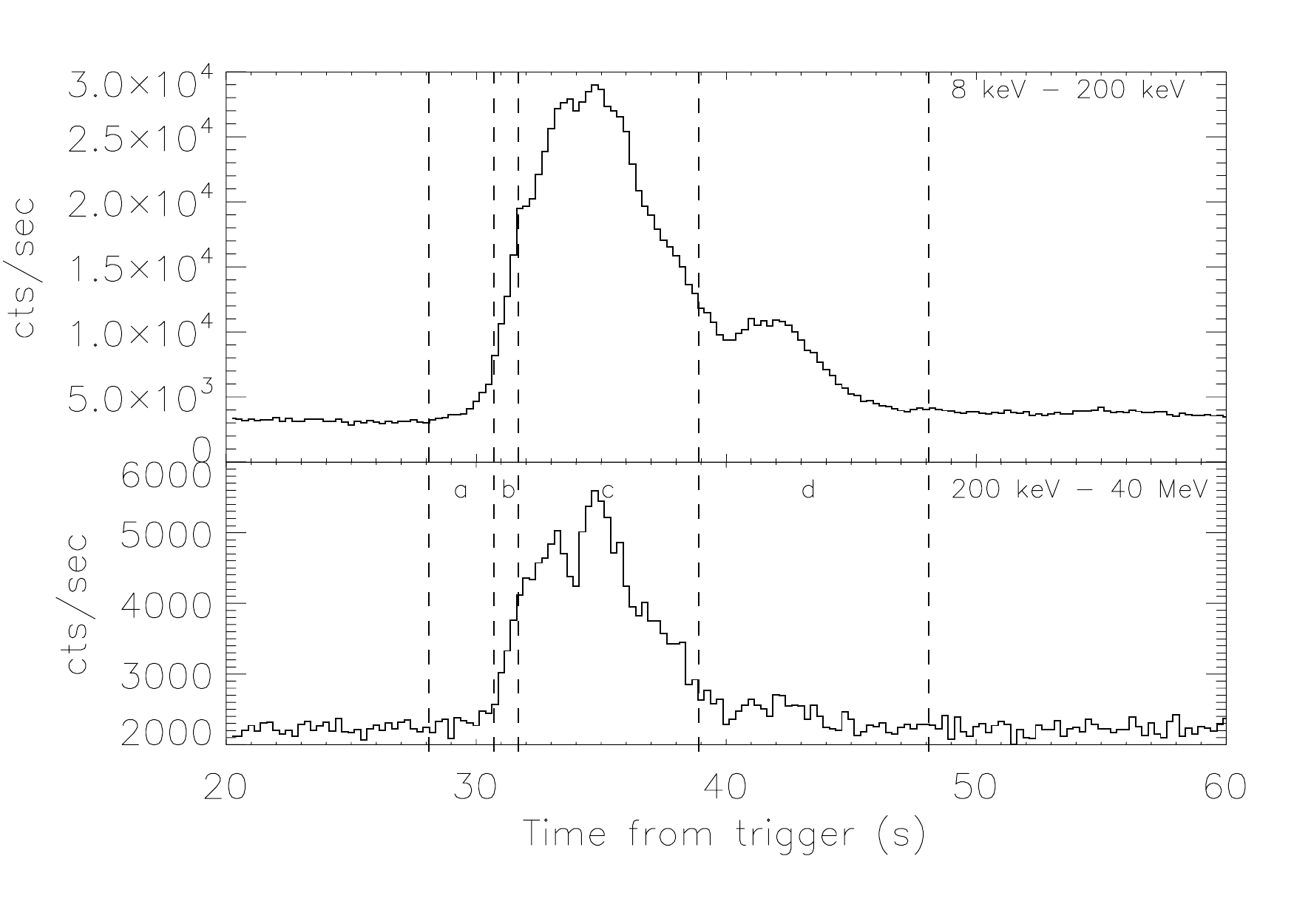}
\caption{ Light curve of GRB~{\it 090820}A as observed by GBM. The two panels show the count rate in the two NAI detectors (top) and BGO (bottom). The dashed lines indicate the time intervals (a, b, c, d) used for the time-resolved analysis (see Figure 3 and Table 1). It is clear that the burst consists of two main peaks and that this burst is very bright in the BGO detectors.}
\label{figure1}
\end{figure}

\begin{figure}
\includegraphics[scale=.6, angle=0]{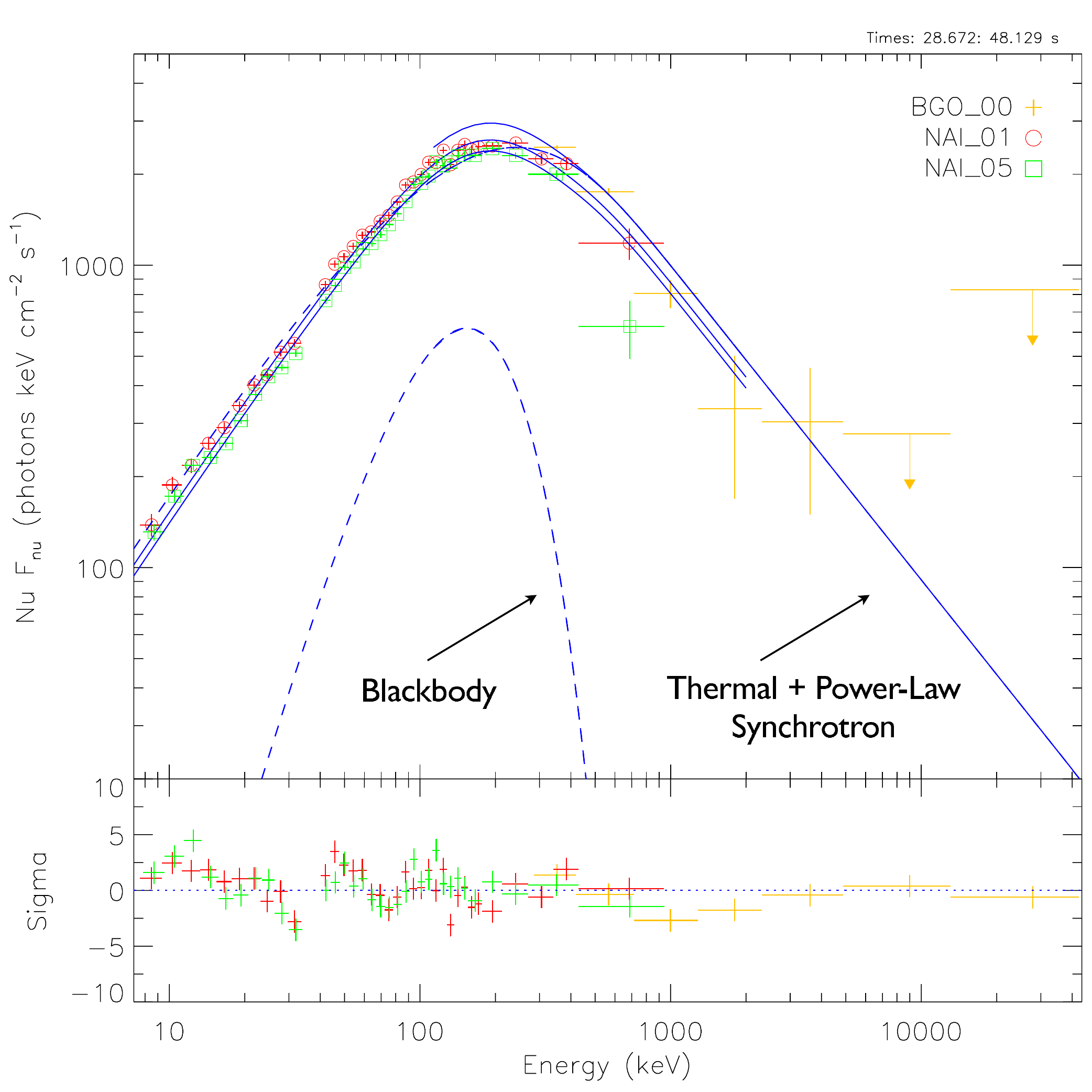}
\caption{The integrated spectrum of GRB~{\it 090820}A. We are able to resolve three components, thermal synchrotron, power-law synchrotron, and a blackbody. Energy channels near the NaI K-edge are omitted. The deviations in the fit residuals are the due to systematics in the detector response resulting from the high count rate and spectral hardness of this burst. However, deviations are never greater than 4$\sigma$ and do not significantly impact the values of the best fit parameters. The multiple curves near the peak of the spectrum are an artifact of the effective-area correction applied to each detector and not related to the different fitted models.}
\label{figure2}
\end{figure}


\begin{figure}
\includegraphics[scale=.5, angle=0]{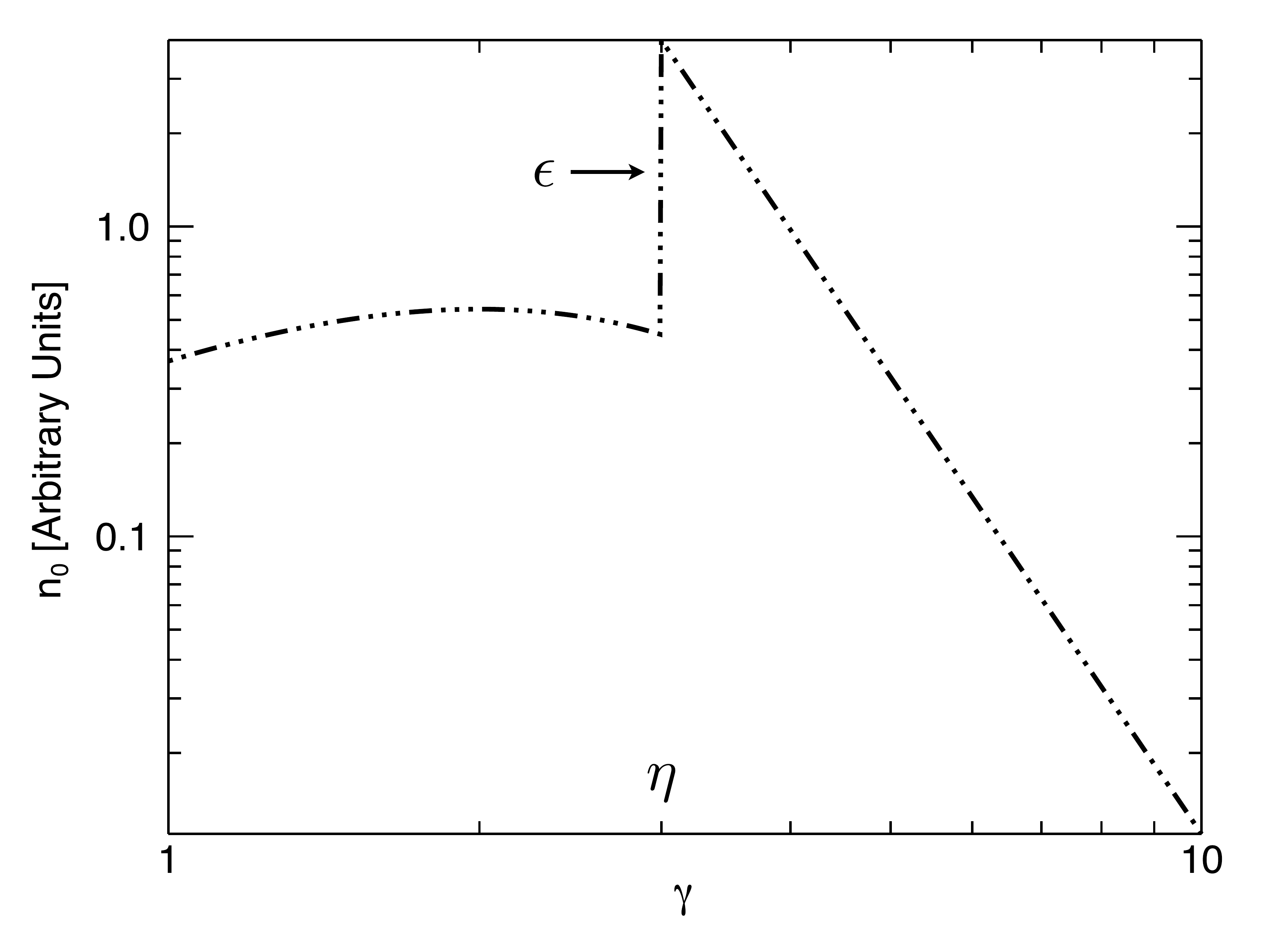}
\label{figure3}
\caption{The electron distribution corresponding to the integrated spectrum. The non-physical jump in the amplitude between the Maxwellian and the power-law distribution (parametrized by $\epsilon$) at $\eta$ is clearly seen.}
\end{figure}

\begin{figure}
\includegraphics[scale=.5, angle=0]{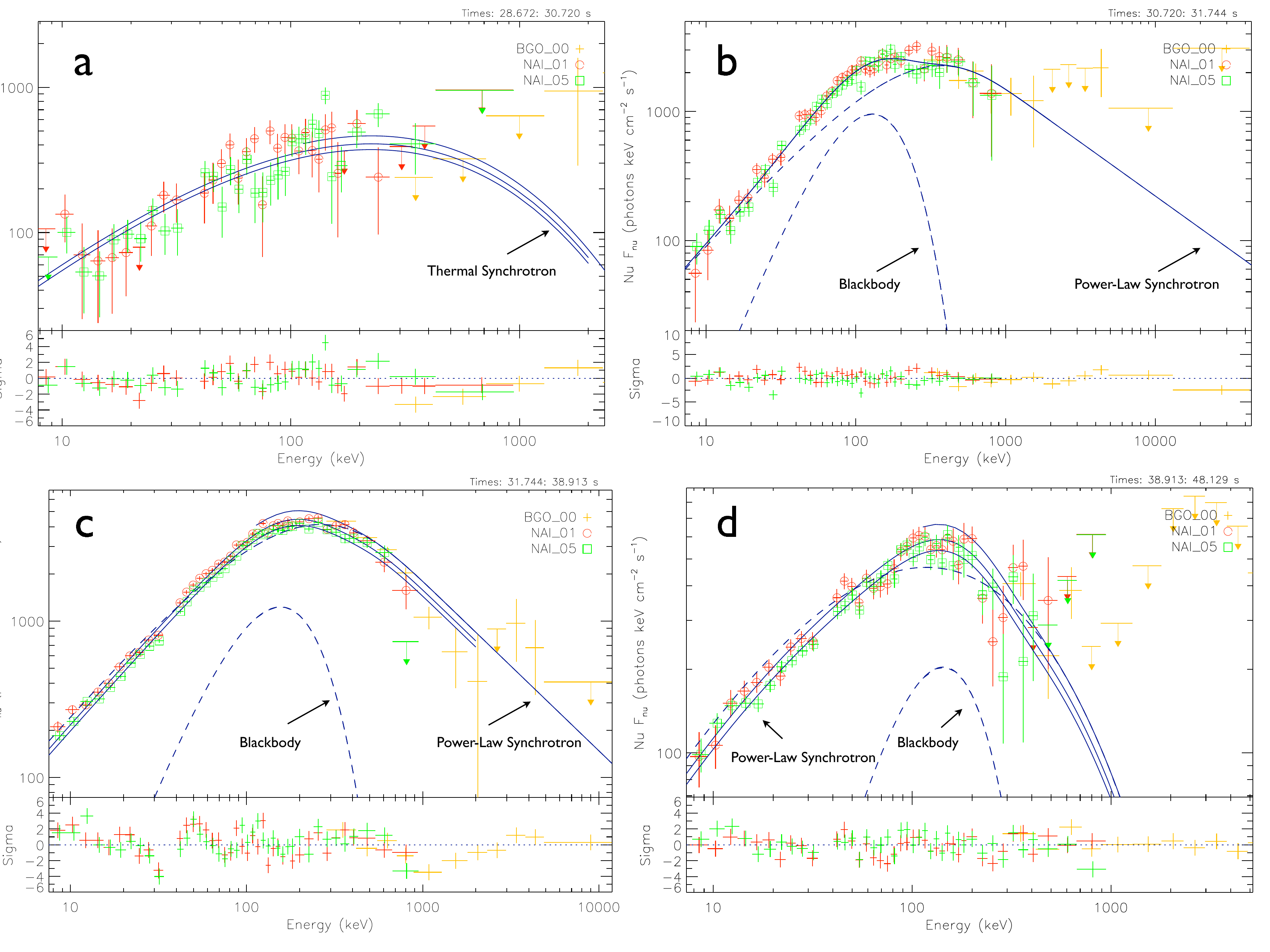}
\label{figure4}
\caption{The time-resolved spectra for GRB~{\it 090820}A. The spectra represent bin \textbf{a} with thermal synchrotron only (top left panel), bin \textbf{b} with power-law synchrotron + blackbody (top right panel), bin \textbf{c} again with power-law synchrotron + blackbody (bottom left panel), and finally bin \textbf{d} with thermal synchrotron + blackbody (bottom right panel). As with Fig. \ref{figure2}, the multiple curves are associated with the effective area correction.}
\end{figure}

\begin{figure}
\includegraphics[scale=.5, angle=0]{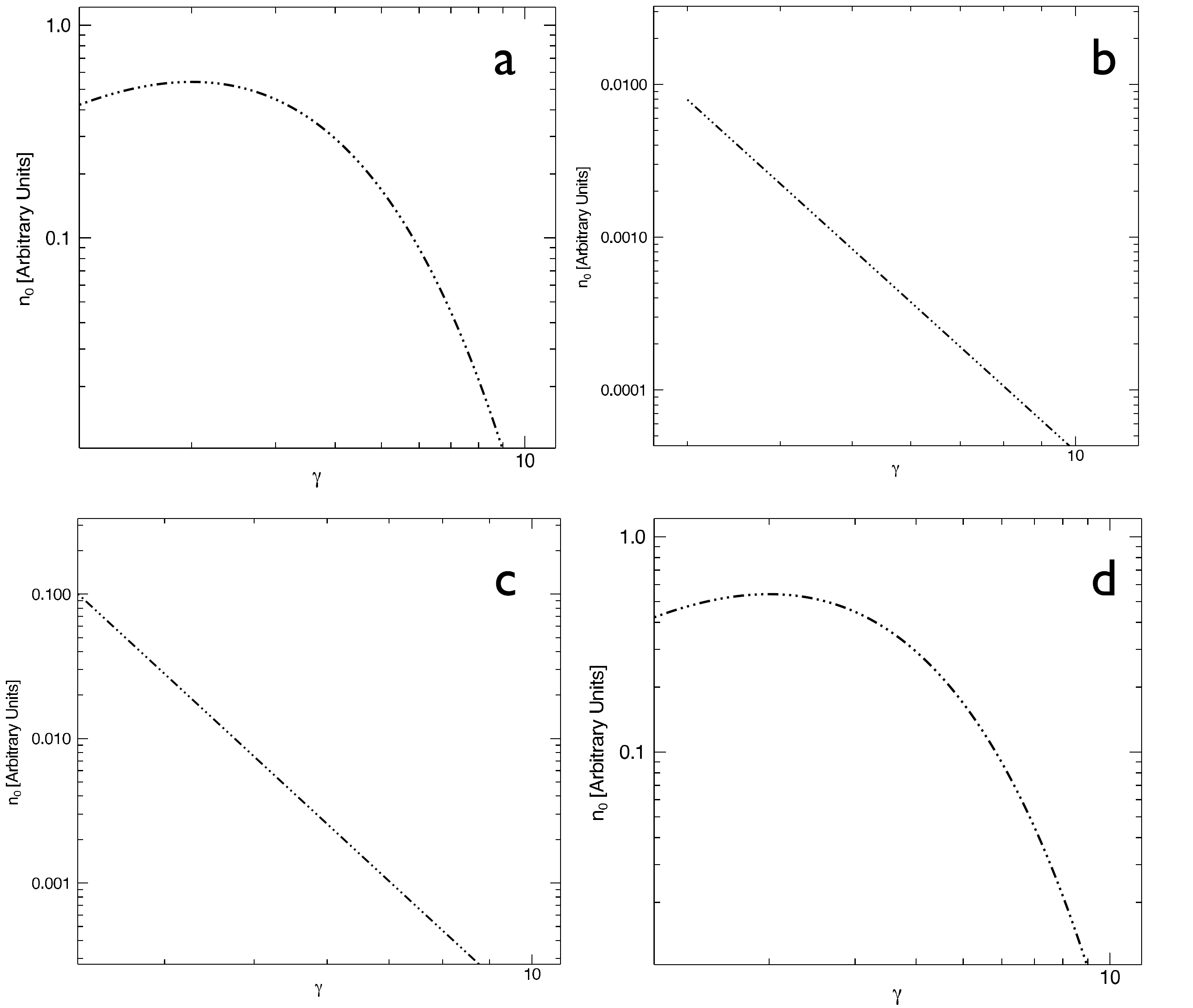}
\label{figure5}
\caption{The electron distributions for the time-resolved spectra. The choice of $\eta$ with a power-law only distribution is arbitrary due to the fact that $\mathcal{E}_c$ and $\eta$ both scale $E_{peak}$.}
\end{figure}



\begin{deluxetable}{c c c c c c c c c}

\renewcommand{\thefootnote}{\alph{footnote}}
\tablecolumns{8}
\tablewidth{0pc}
\tablecaption{The fit parameters for the time-integrated (first row) and time-resolved spectra. The fit parameters for the blackbody component are its amplitude ($A_{BB}$) and energy ($kT$). The fit parameters for the non-thermal components are described in section \ref{sec:model}. The break energy $\mathcal{E}_b\equiv \mathcal{E}_c(\gamma \to \eta\gammat )$ corresponds 
to employing the substitution $\gamma \to \eta\gammat $ in Eq.~(\ref{eq:ergcsynch}).
 Note that the ratio of the amplitudes is not equal to the ratio of the fluxes.}
\startdata
Time interval&Model&$n_{0}$ {$(\gamma s^{-1}cm^{-2}keV^{-1})$} & $\epsilon$ & $\mathcal{E}_c$ (keV) & $\delta$ & $\eta$ & $A_{BB}$ $(\gamma s^{-1}cm^{-2}keV^{-1})$ & $kT$ (keV) \\
\hline \hline
Time integrated&TS+PLS+BB&$0.3437_{-0.065}^{+0.204}$ & $871_{-234}^{+254}$ & $10.39_{-0.245}^{+0.254}$ & $4.9\tablenotemark{a}$ & $3.0$ & $2.08_{-0.208}^{+0.367}\times10^{-5}$ & $42.27_{-1.35}^{+1.49}$ \\

a&TS&$ 2.378_{-0.176}^{+0.189}$ & $-$ &   $8.351_{-0.93}^{+1.08}$ & $-$ & $-$ & $-$ & $-$ \\

b&PLS+BB&$ 859_{-89.1}^{+94.0}$ & $-$ & $14.24_{-0.776}^{+0.848} $ & $ 4.4\tablenotemark{a} $ & $3.0$ & $1.774_{-0.356}^{+0.410}\times10^{-4}$ & $35.32_{-1.77}^{+1.99}$ \\

c&PLS+BB& $1.901_{-0.093}^{+0.094}\times10^{4}$ & $-$ & $15.22_{-0.399}^{+0.411}$ & $ 5.9\tablenotemark{a} $ & $3.0$ & $1.818_{-0.344}^{+0.400}\times10^{-4}$ & $ 38.7_{-1.92}^{+2.13}$ \\

d&TS+BB&$ 2.196_{-0.466}^{+0.720}$ & $-$ & $4.035_{-0.715}^{+0.689}$ & $-$ & $-$ & $8.383_{-3.18}^{+4.89}\times10^{-5}$ & $ 28.40_{-3.59}^{+3.73}$ \\
\enddata
\tablenotetext{a}{fixed at best-fit value}
\label{table1}

\end{deluxetable}


\begin{table}

\begin{tabular}{c|c c c c c}
Time Interval & Band & TS & TS + BB & PLS & PLS + BB \\
\hline \hline
a & 464/355 & 466/357 & 464/355 & 467/357 & 465/355 \\

b & 432/355 & 742/357 & 445/355 & 555/357 & 434/355 \\

c & 450/355 & 1088/357 & 488/355 & 558/357 & 434/355 \\

d & 404/355 & 421/357 & 403/355 & 406/357 & 405/355 

\end{tabular}
\caption{The c-stat per degree of freedom for each time model in the selected time intervals.}
\label{table2}
\end{table}


\begin{thebibliography}{}



\bibitem[Band et~al.(1993) ]{Band:1993p20784}
Band, D., Matteson, J., Ford, L., \& Schaefer, B. 1993,\apj,\vol{413}{281}

\bibitem[Baring(2011)]{Baring:2010p26599}
Baring, M.~G. {2011}, \asr, \vol{47}{1427}

\bibitem[Baring \& Braby(2004)]{Baring:2004p10999}
Baring, M.~G. \& Braby, M. 2004, \apj,\vol{613}{460}

\bibitem[Briggs et al.(2011)]{briggs_ann}  
Briggs, M.~S. et al. 2011 in preparation

\bibitem[Connaughton(2009)]{GCN-GBM}
Connaughton, V. 2009, GCN {Circular} 9829

\bibitem[Chakrabarti et~al.(2009)]{GCN-CORONAS}
Chakrabarti, S., et al. 2009, GCN Circular 9833

\bibitem[Goodman (1986)]{Goodman:1986}
Goodman, J. 1986, \apjl,\vol{308}{L47}


\bibitem[Guiriec et~al.(2010)]{Guiriec:2010p26991}
Guiriec, S., Connaughton, V., Briggs, M., \& Burgess, M., et al. 2010,
\apjl,\vol{727}{L33}


\bibitem[Lloyd \& Petrosian(2000)]{Lloyd:2000p23470}
Lloyd, N. \& Petrosian, V. 2000, \apj,\vol{543}{722}

\bibitem[Lloyd-Ronning \& Petrosian(2002)]{LloydRonning:2002p17917}
Lloyd-Ronning, N. \& Petrosian, V. 2002, \apj,\vol{565}{182}

\bibitem[Meegan et~al.(2009)]{Meegan:2009p28139}
Meegan, C., et al. 2009, \apj,\vol{702}{791}

\bibitem[M{\'e}sz{\'a}ros(2001)]{Meszaros:2001p28197}
M{\'e}sz{\'a}ros, P. 2001, Science, 291, 79

\bibitem[M{\'e}sz{\'a}ros (2002)]{Meszaros:2002}
 M{\'e}sz{\'a}ros, P. 2002, AR \& A, \vol{40}{137}
 
 \bibitem[Niemiec(2004)]{NO04}
Niemiec, J., \& Ostrowski, M. 2004, \apj,\vol{610}{851}

\bibitem[Piran(1999)]{Piran:1999p12227}
Piran, T. 1999, Phys. Rep. \vol{314}{575}

\bibitem[Preece et~al.(2002)]{Preece:2002p11128}
Preece, R., Briggs, M.~S., Giblin, T., \& Mallozzi, R. 2002, \apj,\vol{581}{1248}

\bibitem[Preece et~al.(1998)]{Preece:1998p11011}
Preece, R., et al. \& Pendleton, G.~N. 1998, \apjl,\vol{506}{L23}

\bibitem[Rees \& M{\'e}sz{\'a}ros(2005)]{Rees:2005}
Rees, M.J. and M{\'e}sz{\'a}ros, P. 2005, \apj,\vol{506}{L23}

\bibitem[Rybicki \& Lightman(1979)]{rybicki}
Rybicki, G. \& Lightman, A. 1979, Radiative Processes in Astrophysics (New
  York, Wiley and Sons)
  
\bibitem[Ryde and Pe'er (2009)]{Ryde:2009}
Ryde, F. and Pe'er, A. 2009, \apj,\vol{702}{1211}

\bibitem[Ryde et~al.(2010)]{Ryde:2010p21261}
Ryde, F., et al. , A.~P. 2010, \apjl,\vol{709}{L172}

\bibitem[Spitkovsky (2008)]{Spitkovsky08}
Spitkovsky, A. 2008, \apj,\vol{682}{L5}

\bibitem[Tavani(1996)]{Tavani:1996p23809}
Tavani, M. 1996, \prl\vol{76}{3478}

\end{thebibliography}
\end{document}